# Lunar Cold Trap Contamination by Landing Vehicles

S.T. Shipley[1], P.T. Metzger[2] and J.E. Lane[3]

[1]Enterprise Advisory Services, Inc., NASA ESC-57, Kennedy Space Center, FL 32899, Ph (321) 867-7035, email: scott.t.shipley@nasa.gov
[2]NASA, Granular Mechanics and Regolith Operations Lab, NE-S, Kennedy Space Center, FL 32899, Ph (321) 867-6052, email: phillip.t.metzger@nasa.gov
[3]Enterprise Advisory Services, Inc., NASA ESC-58, Kennedy Space Center, FL 32899, Ph (321) 867-6939, email: john.e.lane@nasa.gov

## ABSTRACT

Tools have been developed to model and simulate the effects of lunar landing vehicles on the lunar environment (Metzger, 2011), mostly addressing the effects of regolith erosion by rocket plumes and the fate of the ejected lunar soil particles (Metzger, 2010). These tools are being applied at KSC to predict ejecta from the upcoming Google Lunar X-Prize Landers and how they may damage the historic Apollo landing sites. The emerging interest in lunar mining poses a threat of contamination to pristine craters at the lunar poles, which act as "cold traps" for water and may harbor other valuable minerals Crider and Vondrak (2002). The KSC Granular Mechanics and Regolith Operations Lab tools have been expanded to address the probability for contamination of these pristine "cold trap" craters.

The KSC tools include simulations of 3D flux of rocket plume ejecta. The trajectories of such ejecta can be mapped onto cold trap craters to predict deposition for expected lunar landings, assuming that the collection efficiency of the 40K cold trap surfaces is 100%. The processes addressed by the KSC tools have now been expanded to address volatile contamination of the lunar surface (Stern, 1999). Landing nearby such a crater will result in the migration of significant exhaust plume gas into the cold trap of the crater, and will also create an unnatural atmosphere over the volatile reservoirs that are to be studied. Our calculations address: 1) the time for the plume-induced local atmosphere above cold traps to decay to normal levels, 2) the efficiency of gas migration into a permanently shadowed crater when the landing is outside it but nearby, and 3) reduction on contamination afforded by moving the landing site further from the crater or by topographically shielding the crater from the direct flux of a lander's ground jet. We also address plume volatiles adsorbed onto and driven inside soil ejecta particles from their residence in the high pressure stagnation region of the engine exhaust plume, and how their mechanical dispersal across the lunar surface contributes to the induced atmosphere. One additional question is whether the collection of soil ejecta along the base of a topographic feature will produce a measurable plume volatile release distinct from the background. We mostly address item 2). Item 3) is obvious from our results excepting that the removal distances may be large, but changes to landing strategy can improve the situation.

INTRODUCTION

The fate of small particles and volatiles (e.g. molecules) emitted from the lunar surface are estimated assuming ballistic trajectories following Keplerian orbits. Under ideal conditions, we find that such particles either escape from the lunar exosphere, or collide with the lunar surface within <u>one orbit</u> about the lunar center of mass. Following Stern (1999), those ideal small particles which collide with the lunar surface are assumed to be reattached to the surface until they are "re-emitted" into the lunar exosphere by either a) thermal desorption or b) sputtering by solar photons or charged particles striking the lunar surface.

The particle velocity distribution is a function of temperature and particle mass, and we assume the emitted particles follow the Maxwell-Boltzmann velocity distribution. Using the Jean's Escape Velocity for lunar surface, the fraction of particles which escape the lunar exosphere is dependent on the scale parameter of the particle ensemble $(kT/m)^{1/2}$ or $(2RT/M)^{1/2}$, where T is kinetic temperature [K], m is particle mass [kg], M is molar mass [g/mole], and k & R are the appropriate ideal gas constants for the chosen units. Given additional information on the exhaust rate of release [kg/sec] and total mass of the emitted particles, we calculate the time dependent concentration of particles in the lunar exosphere, and characteristic time for deposition of the "emitted" particles to the lunar surface. Such "in-flight" concentrations are needed to estimate the rate of particle loss due to exospheric collisions with solar UV radiation or the solar wind plasma. These calculations are made for ideal conditions assuming a spherical Moon with no surface terrain, a spherical gravity field with all mass concentrated at the Lunar Center, and no aspherical/asymmetric gravitational effects. The lunar exosphere is considered to be sufficiently rarified so that there are no collisions after each particle has been emitted from the lunar surface. The effects of electric fields are also neglected in this paper, so we are addressing only the propagation of neutral (uncharged) particles.

**BALLISTIC TRAJECTORIES IN THE LUNAR EXOSPHERE**

Consider a particle or other small mass released from the lunar surface into the lunar exosphere, with no limit to its mean free path. Assuming a spherical moon with all mass concentrated at its center of mass and no forces other than gravity, this particle will follow a Keplerian elliptical orbit. The orbital equations provide the distance **r** of such particles from the center of the larger mass (not center of ellipse)

$$r = \rho (1 + \varepsilon \cos \theta)^{-1} \qquad \text{Eq. (1)}$$

where ρ is some constant (defined later), and

ε     eccentricity, $\varepsilon = (1 - b^2/a^2)^{1/2}$
θ     True Anomaly
a     semi-major axis of ellipse (defined along apoapsis)
b     semi-minor axis of ellipse.

Assuming that the particle will reach its highest altitude above the lunar surface at the orbital apoapsis along the semi-major axis of the elliptical trajectory, then each orbit can be determined from initial conditions $v_o$ and $\varphi_o$ at point **p** as shown in Figure 1.

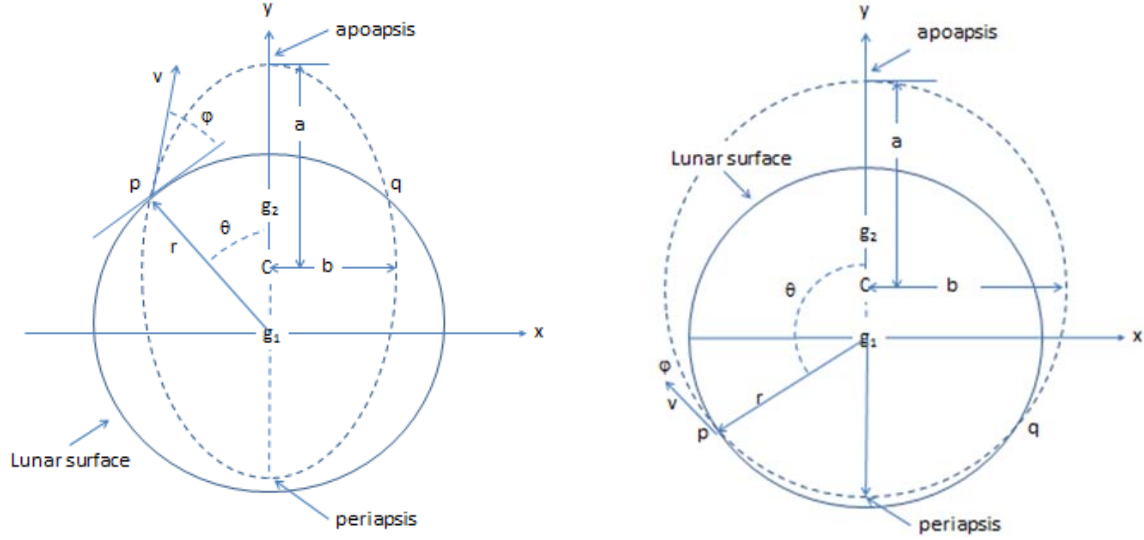

**Figure 1. Elliptical orbital trajectories for particle emitted from lunar surface at point p, reaching highest altitude above lunar surface along the semi-major axis (apoapsis), and returning to the lunar surface at point q.**

Given initial conditions for particle speed $v_o$ and direction $\varphi_o$ at the lunar surface, then the normalized angular momentum **h** for any particle is conserved for a lossless orbit such that

$$h = r_o \cdot v_o \cdot \cos \varphi_o$$

where **h** is constant. Note that $r_o = r_{moon}$ for initial conditions at lunar surface with $r_{moon}$ being the radius of an ideal spherical Moon. Angular momentum $h \cdot m$ is also conserved since particle mass **m** is constant. Following Vinti (1998), the orbit equation defines normalized total energy:

$$W = \tfrac{1}{2} v^2 - \mu / r \;<\; 0$$

where $\mu = GM$, G is the Universal Gravitation Constant and M is Total Lunar Mass. The semi-major axis a, orbit constant $\rho$ and eccentricity $\varepsilon$ are

$$a = -\mu / 2W$$
$$\rho = h^2 / \mu$$
$$\varepsilon = (1 - \rho / a)^{1/2} = (1 - v^2 r^2 \cos^2 \varphi / \mu a)^{1/2}$$

where Jean's Escape Velocity for particles leaving the lunar surface and $W = 0$ is

$$v_{esc} = (2\mu / r_{Moon})^{1/2} \qquad\qquad \text{Eq. (2)}$$

Simplifying by substitution of $v_{esc}$ from Eq. 2, then given $v_o$ and $\varphi_o$ at $r_o$

$$r_o / a = 2(1 - v_o^2 / v_{esc}^2) \qquad\qquad = f(v_o)$$
$$\varepsilon = [1 - 2(r_o / a) \cdot \cos^2 \varphi_o \cdot v_o^2 / v_{esc}^2]^{1/2} \qquad = f(v_o, \varphi_o)$$
$$\theta = \arccos[\varepsilon^{-1}(2\cos^2 \varphi_o \cdot v_o^2 / v_{esc}^2 - 1)] \quad = f(v_o, \varphi_o) \qquad \text{Eq. (3)}$$

The length between focus $g_1$ and geometric center C in Figure 1 is $a\varepsilon$, and the height $z_a$ of the particle above the lunar surface at apoapsis is

$$z_a = a(1 + \varepsilon) - r_o \qquad > 0$$

Particles emitted from the lunar surface with no additional collisions will escape the lunar environment if their velocity exceeds $v_{esc}$ = 2.375 km s$^{-1}$. Substituting $u = v_o/v_{esc}$ in Eq. (3), then

$$\rho = 2\, u^2 \cdot \cos^2 \varphi_o$$

where $0 \leq \theta < \pi/2$ for $\rho < 1$, $\theta = \pi/2$ for $\rho = 1$, and $\pi/2 < \theta \leq \pi$ for $\rho > 1$. True Anomaly is shown as a function of u and $\varphi_o$ in Figure 2. This figure shows which fraction of the particles are lost to gravitational escape ($u \geq 1$), which land in the hemisphere of point **p** ($\rho < 1$) and which land in the opposing hemisphere ($\rho > 1$). Orbit eccentricity is shown in Figure 3. The trivial case where $\varphi_o = 0$ and $u_o = 2^{-\frac{1}{2}}$ defines a circular orbit following an ideal path along the surface of a spherical Moon. Although theoretically possible, such trajectories are not likely due to terrain interception, and orbital variations due to asymmetric gravitational fields.

We developed an Excel spreadsheet and FORTRAN algorithms to calculate semi-major axis **a** [km], eccentricity ε, height above lunar surface at apoapsis $z_a$ [km], Great Circle path $2\theta$ [radians], and time in flight **t** [sec] between initial emission point **p** and final deposition point **q**. Sample results are shown in Figure 4. The heights of the orbital apoapsis Above Lunar Surface (ALS) are shown in Figure 5.

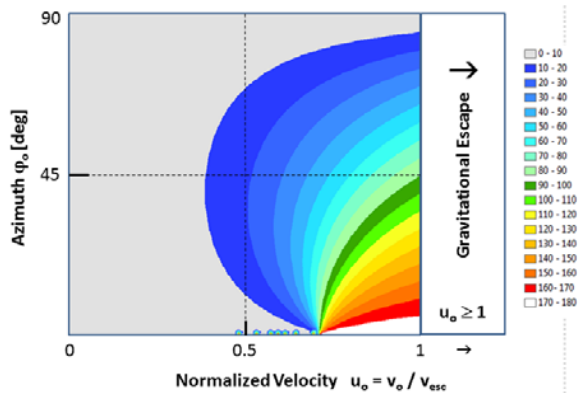

**Figure 2. The velocity-elevation partition of True Anomaly $\theta$ for particles emitted from point p on the lunar surface.**

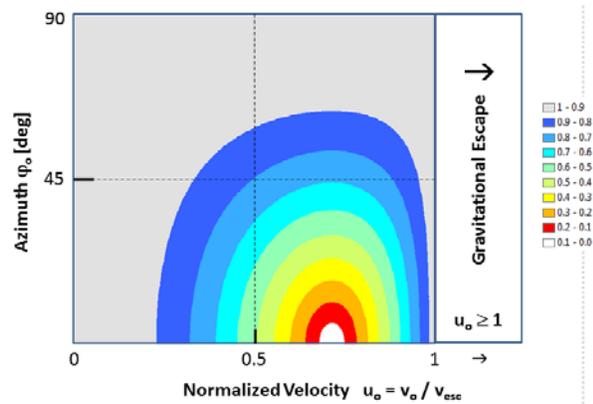

**Figure 3. Ballistic orbit eccentricity as a function of normalized velocity $u_o$ and azimuth angle $\varphi_o$.**

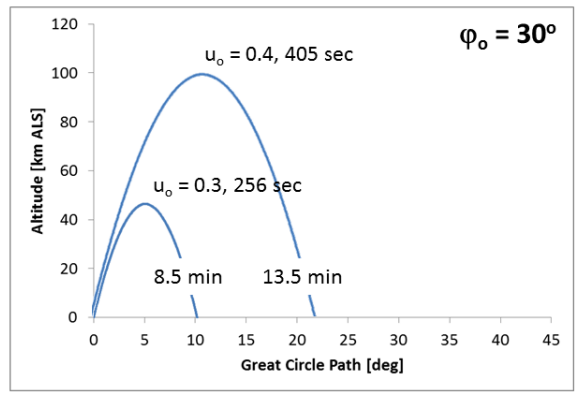

**Figure 4. Particle trajectories assuming ballistic elliptic orbits for $\varphi_o = 30°$ and selected initial velocities.**

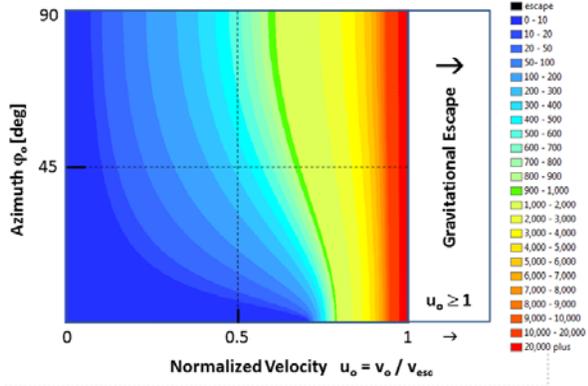

**Figure 5. Apoapsis height in meters Above Lunar Surface (ALS) as a function of normalized velocity $u_o$ and azimuth angle $\varphi_o$.**

# THE MAXWELL-BOLTZMANN VELOCITY DISTRIBUTION

Assuming a Maxwell distribution, then the statistical ensemble of particle velocities will be a function of kinetic temperature T [K] and particle mass m [kg], such that:

$$B(v) = 4/\sqrt{\pi} \cdot v^2 / v_p^2 \cdot \exp(-v^2 / v_p^2)$$

where k is the Boltzmann constant [kg m$^2$ s$^{-2}$ K$^{-1}$]. The peak value or "characteristic scale parameter" for this distribution is

$$v_p = (2 k T / m)^{1/2} = (2 R T / M)^{1/2}$$

for ideal gas constant R = 8314 m$^2$ s$^{-1}$ K$^{-1}$ and molecular mass for the ideal gas M [g mole$^{-1}$]. The list of mostly neutral particles is derived from the expected products of a popular hypergolic fuel, which can be found in Table 1. The fraction of particles exceeding the lunar escape velocity $v_{esc}$ for characteristic molecules and temperatures is given in Table 2.

Table 1 – Characteristic Exhaust Products of a Common Hypergolic Engine Propellant

| Propellant A | Propellant B (Oxidizer) | Exhaust Products (Haas, 1984) |
|---|---|---|
| Aerozine 50 (50% UDMH, (CH$_3$)$_2$N(NH$_2$) 50% Hydrazine, N$_2$H$_4$) | Dinitrogen Tetroxide (NTO), N$_2$O$_4$ | H2O, N2, CO, CO2, H2, OH *Note: relative concentrations of combustion products appears to depend on fuel reaction temperatures.* |

Given molar mass M and kinetic temperature T, the scale velocity parameter $v_p$ is used to calculate the velocity probability distribution B(v). The cumulative probability of B(v) is used to estimate the fraction of the distribution exceeding a characteristic velocity such as the Jean's Escape value for surface emissions. Sample B(v) distributions are shown in Figures 6 and 7.

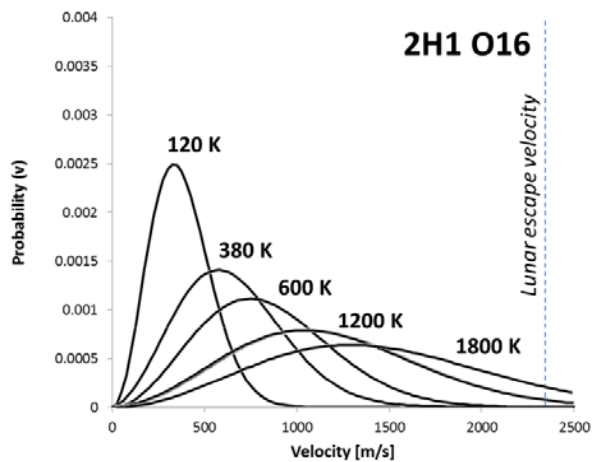
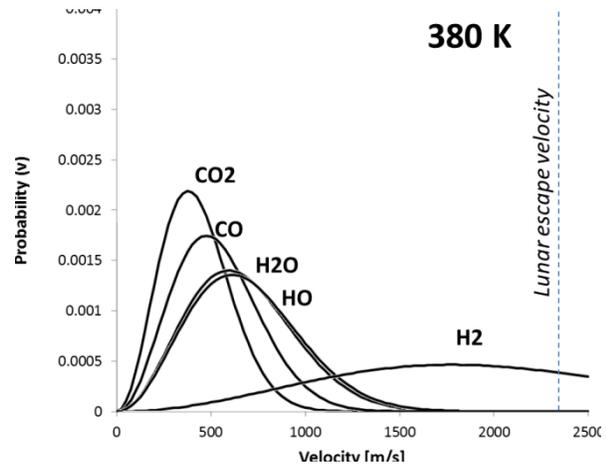

Figure 6. Maxwell Boltzmann distribution for 2H1 O16 (18 g mole$^{-1}$) velocity at various temperatures, including daytime lunar surface (380K). The lunar escape velocity is shown as the dashed line to the right.

Figure 7. Maxwell Boltzmann distribution for velocities of Table 1 exhaust products at daytime lunar surface temperature (380K). The lunar escape velocity is shown as the dashed line to the right.

Table 2. Probability of escape for selected volatiles in Lunar Exosphere

| particle | M [g mole$^{-1}$] | T [K] | $v_p$ [m s$^{-1}$] | Prob (v > $v_{esc}$) |
|---|---|---|---|---|
| 2H1 (H$_2$) | 2 | 120 K | 998.8 | 9.62E-01 % |
| | | 380 K | 1,777 | 30.7 % |
| | | 600 K | 2,234 | 51.5 % |
| | | 1200 K | 3,159 | 76.7 % |
| | | 1800 K | 3,868 | 85.9 % |
| H1 O16$^-$ (OH) | 17 | 120 K | 342.6 | none |
| | | 380 K | 609.7 | 9.93E-05 % |
| | | 600 K | 766.1 | 2.23E-02 % |
| | | 1200 K | 1,083 | 2.12 % |
| | | 1800 K | 1,327 | 9.06 % |
| 2H1 O16 (H$_2$O) | 18 | 120 K | 333.0 | none |
| | | 380 K | 592.5 | 4.14E-05 % |
| | | 600 K | 744.5 | 1.29E-02 % |
| | | 1200 K | 1,052 | 1.63 % |
| | | 1800 K | 1,289 | 7.66 % |
| 2N14 or C12 O16 (N$_2$ or CO) | 28 | 120 K | 267.0 | none |
| | | 380 K | 475.0 | 6.13E-09 |
| | | 600 K | 596.9 | 5.22E-05 % |
| | | 1200 K | 844.2 | 1.13E-01 % |
| | | 1800 K | 1,034 | 1.37 % |
| C12 2O16 (CO$_2$) | 44 | 120 K | 213.0 | none |
| | | 380 K | 379.0 | none |
| | | 600 K | 476.2 | 6.90E-09 % |
| | | 1200 K | 673.4 | 1.44E-03 % |
| | | 1800 K | 824.8 | 7.91E-02 % |

**TIME OF FLIGHT FOR LUNAR EJECTA**

We derived two time-of-flight formulations for particles leaving the Lunar surface with velocities $v_o$ not exceeding lunar escape velocity $v_{esc}$, traveling along orbits with constant total energy, and returning to the lunar surface without collisions or other interactions along the way. The geometry of such orbits is shown in Figures 8 and 9. Kepler's Second Law is applied, wherein the area dA/dt swept out by the vector **r** over the True Anomaly θ in equal times is a constant. The total area of an ellipse with semi-major axis **a** and semi-minor axis **b** is πab. This total area corresponds to the orbital time period for a particle returning through periapsis to point **p**:

$$T = \pi \, T_o \cdot (1 - u_o^2)^{-3/2}$$

where $T_o = r_o / v_{esc} \sim 12.16$ min, and $u_o = v_o/v_{esc}$ is a normalized particle velocity. Using fractional area f = 2(A$_1$ + A$_2$)/πab as shown in Figure 8, then the travel time from **p** to **q** is:

$$T_{pq} = f \, T_o \cdot (1 - u_o^2)^{-3/2} \qquad ; 0 \leqq f \leqq \pi$$

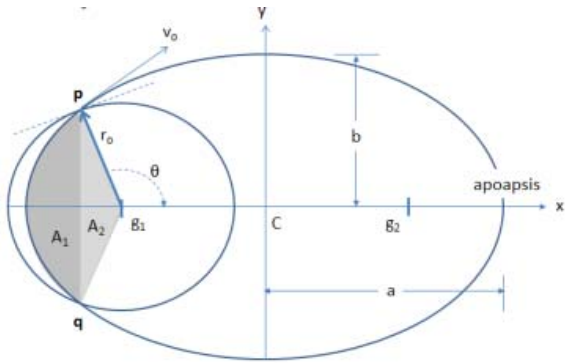 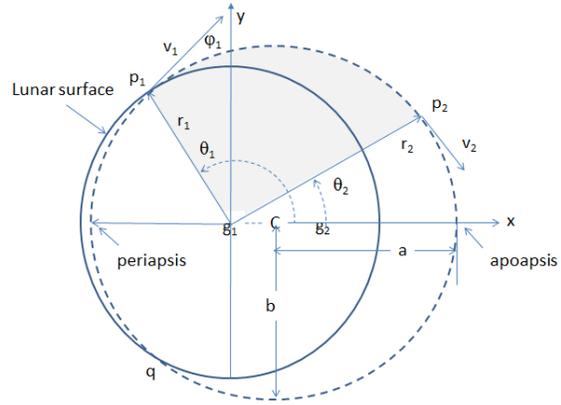

**Figure 8 – Ideal lunar orbital trajectory for particle leaving lunar surface at point p, and moving through the apoapsis and returning to the lunar surface at point q.**

**Figure 9 – Geometry for particle traveling along a partial orbit from $p_1$ to $p_2$. When vectors $r_1$ and $r_2$ are known, then a family of orbits will satisfy $\theta_2 - \theta_1 =$ const.**

where
$$f = \arcsin(s) + s(1-s^2)^{1/2} - r_o^2/ab \cdot \cos\theta \sin\theta + \tfrac{1}{2}\pi \qquad \text{Eq. (4)}$$
$$s = r_o/a \cdot \cos\theta - |\varepsilon|$$

and $-1 \leq s \leq 1$. Note that $f = \pi$ when $\theta = \pi$ and $s = -1$ (full orbit from periapsis), and $f = 0$ when $\theta = 0$ and $s = 1$ (short hop at apoapsis). We also derive the time of flight using the orbit equation (Eq. 1) using the geometry shown in Figure 9, which has the following solution

$$f = \pi - \frac{\rho^2}{ab}\int_\theta^\pi \frac{d\theta}{(1+\varepsilon\cos\theta)^2} \qquad \text{with} \quad \frac{\rho^2}{ab} = b^3/a^3$$

Integrating from $\theta$ to $\pi$, then:

$$f = 2\arctan\left(\frac{\sqrt{1-\varepsilon}}{\sqrt{1+\varepsilon}}\tan\frac{\theta}{2}\right) - 2\varepsilon\sqrt{1-\varepsilon^2}\left(\frac{\sin\theta}{1+\varepsilon\cos\theta}\right) \qquad \text{Eq. (5)}$$

where

$$\lim_{\theta\to\pi} 2\arctan\left(\frac{\sqrt{1-\varepsilon}}{\sqrt{1+\varepsilon}}\tan\frac{\theta}{2}\right) = \pi$$

The time of flight from any True Anomaly $\theta$ to apoapsis is proportional to f and can be derived from either Eq. (4) or Eq. (5). Given $r_1$ and $r_2$ and the Great Circle angular distance $\delta\theta_{12} = \theta_2 - \theta_1$ between them, then there is a family of elliptical orbits which will pass through both $p_1$ and $p_2$. Each of these orbits has a unique eccentricity $\varepsilon$, so the apoapsis height $a(1+|\varepsilon|)$ and the time of flight $f(\varepsilon)$ for each of these orbits are also unique functions of $\varepsilon$. For this analysis, we limit the solutions to $r1 = r2$ and confine the source to the lunar surface.

We solve for $f(\varepsilon)$ given the case were $r_1 = r_2 = r_o$. In such cases the apoapsis of any orbit will be located at $\delta\theta_{12}/2$, such that

$$a = r_o(1-\varepsilon^2)^{-1}\{1 - |\varepsilon|\cos(\delta\theta_{12}/2)\} \qquad ; 0 \leq |\varepsilon| < 1 \quad \text{Eq. (6)}$$

We now have the parameters needed to calculate $f(\varepsilon)$ for each $\varepsilon$, namely $\theta = \delta\theta_{12}/2$ where $0 \leq |\varepsilon| < 1$, a from Eq. (6), and f from Eq. (4) or Eq. (5). The time of flight $f(\varepsilon)$ is therefore a monotonic function of ε, with the shortest time defined by ε = 0 (circular trajectory from $p_1$ to $p_2$). The longest time is obtained as ε approaches 1. Each orbit also defines a unique $\varphi_o$ and $v_o$ at $p_1$. The fate of volatiles in the lunar exosphere can now be obtained by convoluting the time of flight with the velocity distribution as a function of source temperature. We simplify the problem by assuming that emissions from the lunar surface are isotropic (uniform in angle).

Our FORTRAN program EMISSION.F90 integrates the deposition from a point source to Great Circle arc lengths 2θ, summing up all contributions over velocity from 0 to $v_{esc}$ and all emission elevation angles φ from 0 to π/2. Results for H2O gas emissions from a point source are shown in Table 3 for selected gas temperatures. Contours for such "first hop" deposition from a point source are visualized with Google Moon in Figure 10 for water at the characteristic lunar daytime surface temperature of 380 K. It is the inverse problem that we need to solve, namely for each θ find the relative contribution of emissions integrated over the velocity distribution. At this point, such a solution is found numerically, with results shown in Figure 10.

Table 3. H2O gas deposition from point source as function of Great Circle arc [degrees]

| Source Temperature | Fraction lost to space | 50% Deposition after one hop | 67% deposition after one hop | 98% deposition after one hop |
|---|---|---|---|---|
| 120 K | none | 0.7 ° | 1.2 ° | 4.6 ° |
| 380 K | 4.14E-05 % | 2.3 ° | 3.9 ° | 18.7 ° |
| 600 K | 1.29E-02 % | 3.8 ° | 6.7 ° | 39.9 ° |
| 1200 K | 1.63 % | 8.6 ° | 16.0 ° | 139 ° |
| 1800 K | 7.66 % | 13.4 ° | 25.8 ° | 164 ° |

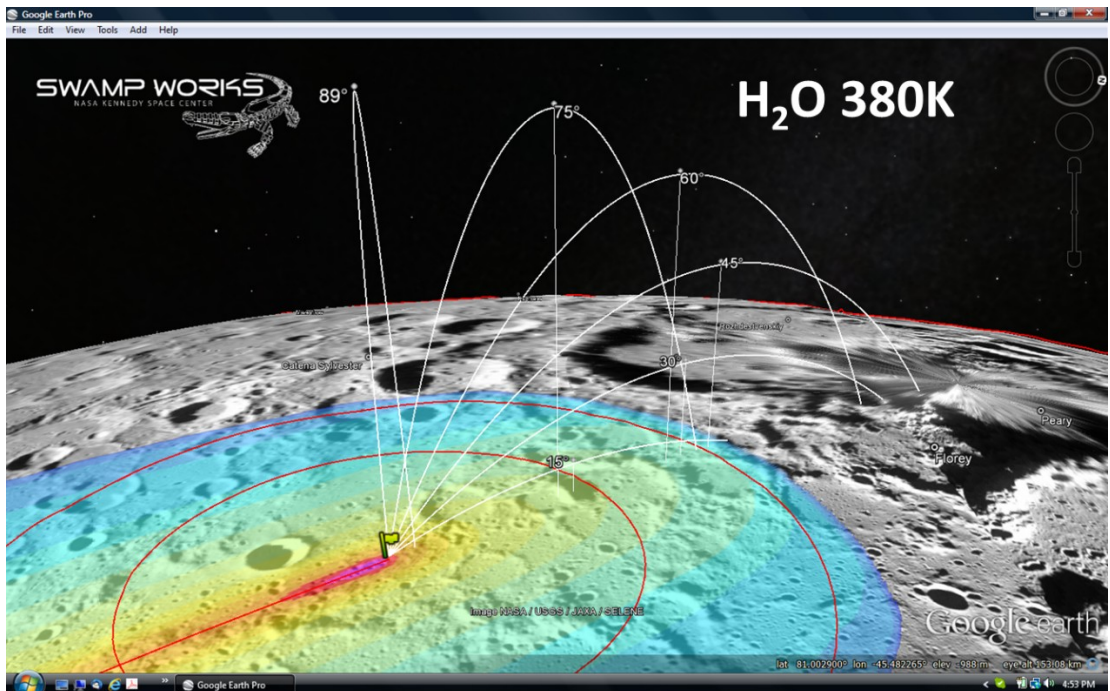

Figure 10. Deposition contours from a point source in Google Moon, using the results for non-ionized H2O gas at 380 K. Trajectories (white) are shown for one hop deposition of H2O volatiles at $v_p$ for 380K, $u_o$ = 0.256. Contours are shown as range rings (red) including 50% and 67% of the deposition, from Table 3.

# GEOMETRY FOR LUNAR DEPOSITION

The deposition of engine exhaust products is estimated for a lunar lander descending to the surface along a ground track **s** with engine height **h** and engine axis tilted down from the local horizontal by angle $\beta_o$. The density of engine exhaust is provided by CFD case studies for an Apollo Lunar Lander (LM), and these results are combined with an ideal spherical geometry model for intersection of rays from the engine exhaust port onto the surface at point **p** defined by along track $s_1$ and cross track $n_1$ coordinates.

**CFD Case Study** – Using CFD simulation for Apollo LM descent engine operating at 45m above the Lunar surface, results have been interpolated to a regular grid by 3-point linear triangulation. Results are shown in Figure 11 for the LM descent engine parameters, with an arbitrary color scheme to highlight the general structure of the derived fields (each color scheme is unique). The combination of density and radial velocity at 25m ALS (Above Lunar Surface) provides an estimate of engine exhaust flux in units of kg m$^2$ sec$^{-1}$ which is shown in Figure 13. This is a compromise, since the 25m ALS height is the lowest altitude in this simulation which provides a cross section of the entire exhaust plume, but may be far enough from the engine to provide a "far field" estimate of that flux. We note that deposition calculations will use distances from the engine on the order of 10 km to 100 km.

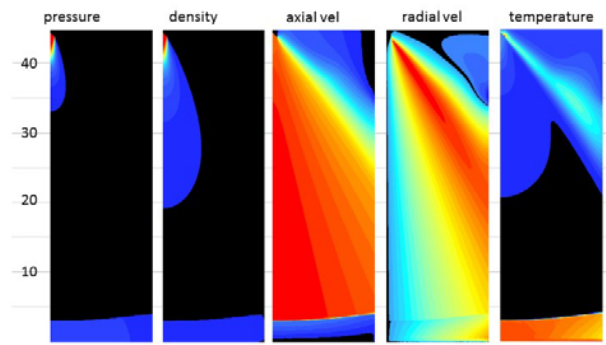

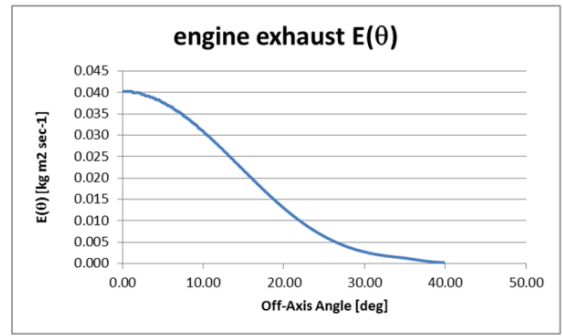

**Figure 11 – CFD Case Study for Apollo LM engine at h = 45m and descending vertically to lunar surface. Density and Axial Velocity at height 25 m ALS are used to estimate the far-field engine exhaust flux.**

**Figure 12. Engine Exhaust flux [kg m$^2$ sec$^{-1}$] as a function of off-axis angle $\theta$ at a distance of 20 m. Approximately 90% of the exhaust is contained within off-axis angle of 21.2°, and 66% within 12.7°.**

**Deposition Geometry** – Consider an engine E located at height h above point $s_o$, tilted downward by angle $\beta_o$ so that the exhaust axis is aligned along ray <u>E E'</u> and oriented above the along-track coordinate s, as shown in Figure 13. Following are equations which estimate the off-axis angle $\theta$ for point **p** on the Lunar surface, using the natural coordinates $\{s_1, n_1\}$ for **p** where **s** is along-track distance [m] and n is cross-track distance [m]. The following derivation is designed to calculate deposition to a rectangular grid in natural coordinates $\{s, n\}$ for an exhaust source located at height h above grid point $\{s_o, 0\}$. Using each point **p** defined by natural coordinates $\{s_1, n_1\}$, the the Great Circle arc $\psi$ from $s_o$ to **p** is

$$\cos \psi = \cos (s_1/R) \cos (n_1/R) \qquad ; \psi \geq 0$$

where R is Lunar radius. Note that $\psi$ defines a range ring about point $s_o$ with radius $s = \psi R$, which is the dashed line **<u>ps'</u>** in Figure 13. Point **p** is at range r from exhaust source E, where

$$r/R = [1 + (1+h/R)^2 - 2(1+h/R)\cos\psi]^{1/2}$$

The tilt of this conical surface downward from horizontal is defined by angle $\beta$, where

$$\cos\beta = (r/R)^{-1} \sin\psi \qquad ; 0 < \beta < \pi/2$$

Note that r defines a cone with apex at source E and intersecting the Lunar surface along arc **ps'** as r is rotated about the z axis through angle $\gamma$, where

$$\sin\gamma = \sin(n_1/R) / \sin\psi \qquad ; 0 < \gamma < \pi$$

The off-axis angle for point p is then

$$\cos\theta = \sin\beta \sin\beta_o + \cos\beta \cos\beta_o \cos\gamma$$

The exhaust flux $E(\theta)$ to point p is derived using the approximation

$$\delta E(\theta) = 16 \text{ kg sr}^{-1} \text{ sec}^{-1} \exp(-8.886 \text{ rad}^{-2} \theta^2) \, \delta\Omega$$

and the solid angle subtended on the surface is

$$\delta\Omega = \delta s \, \delta n \sin\beta / r^2$$

There is a limitation for rays intersecting the lunar surface, which requires that angle $\alpha$ for any point **p** along the arc **ps'** must be greater than $\pi/2$. Otherwise the exhaust particles will miss the surface and either go into orbit or escape the lunar environment. This limitation is defined by

$$\alpha = \pi/2 + \beta - \psi \qquad ; \alpha > \pi/2 \text{ or } \beta > \psi$$

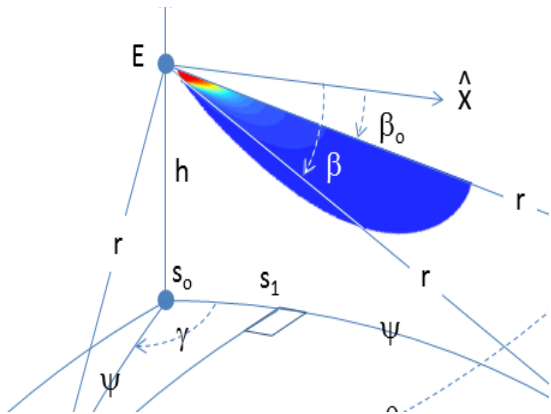

**Figure 13. Spherical geometry for engine exhaust axis tilted downward by angle $\beta_o$ and axially symmetric about ray E E'. The off-axis angle $\theta$ is estimated for point p at natural coordinates $\{s_1, n_1\}$.**

**Deposition for Apollo 17 Lunar Module Descent** – The deposition rate [km m-2 sec-1] to the lunar surface is calculated using the known descent pattern and engine parameters for the Apollo 17 Lunar Module (LM). The instantaneous deposition pattern to the surface at start of braking is shown in Figure 14. The LM will land at 720 sec and 522 km down range after start of braking. Using the LM descent profile shown in Table 4, the instantaneous deposition pattern assuming bursts every 120 sec is calculated and shown in Figure 15. These results are shown combined with the LM descent geometry and exhaust cones using Google Moon in Figure 16. The full deposition map is shown on normal coordinates in Figure 17, and is shown remapped to Google Moon in Figure 18.

Table 4. Apollo 17 LM Descent Parameters (used in Figure 16)

| time | s [km] range from braking | beta [deg] engine tilt | h [km] engine height |
|---|---|---|---|
| 0 sec | 0 | 0 | 17.2 |
| 120 sec | 190 | 12.0 | 14.8 |
| 240 sec | 401 | 12.6 | 12.6 |
| 360 sec | 446 | 22.0 | 9.5 |
| 480 sec | 502 | 28.9 | 6.1 |
| 600 sec | 520 | 59.0 | 0.86 |
| 720 sec | 522 | 90.0 | 0 |

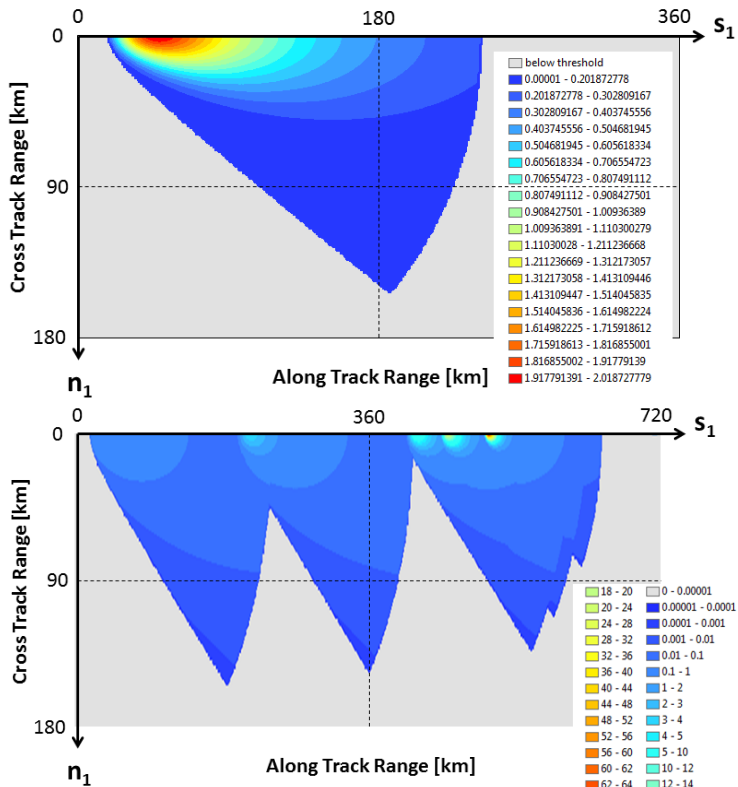

Figure 14. Apollo 17 LM descent engine deposition rate [g km$^{-2}$ sec$^{-1}$] for 1 sec at start of braking at t= 0 sec (LM at upper left). The distance cross track is shown from 0 (top) to 180 km (bottom).

Figure 15. Apollo 17 LM descent engine deposition rate [g km$^{-2}$ sec$^{-1}$] at selected times (every 120 sec) with braking starting at t= 0 sec (upper left). Range shown along track is 720 km, with braking starting at s = 0 km (left) and landing at s = 522 km.

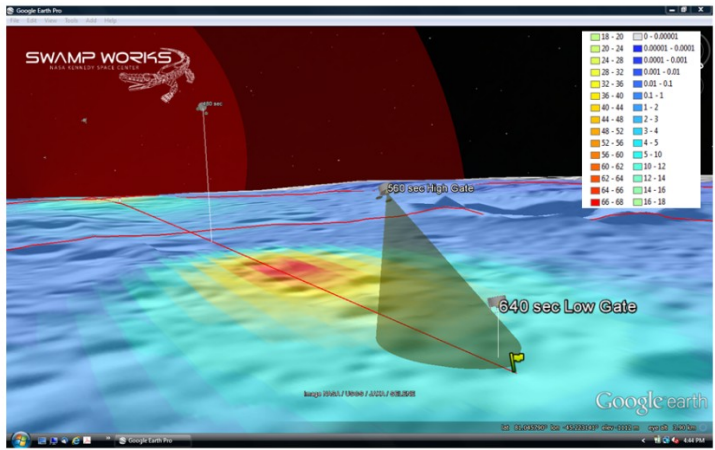

Figure 16. Lunar deposition bursts from Figure 15 remapped onto Google Moon, with LM positions and model of exhaust jet cone.

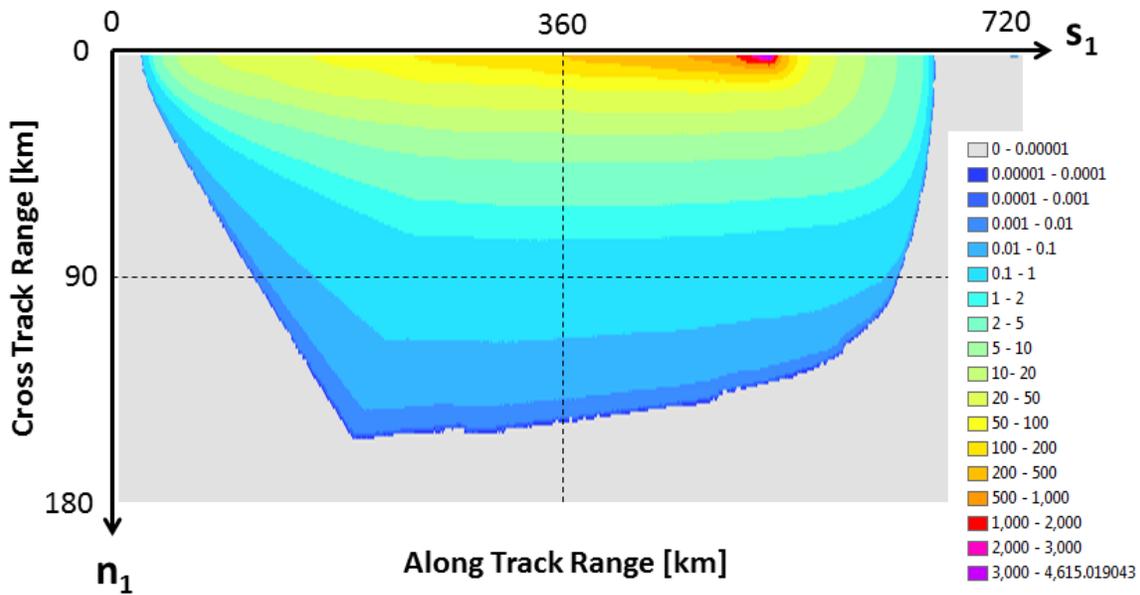

**Figure 17.** Apollo 17 LM descent engine total deposition [g km$^{-2}$] after braking starting at s = 0 km (left) and landing at s = 522 km. Range shown along track is 720 km, The distance cross track is shown from 0 (left) to 180 km (right).

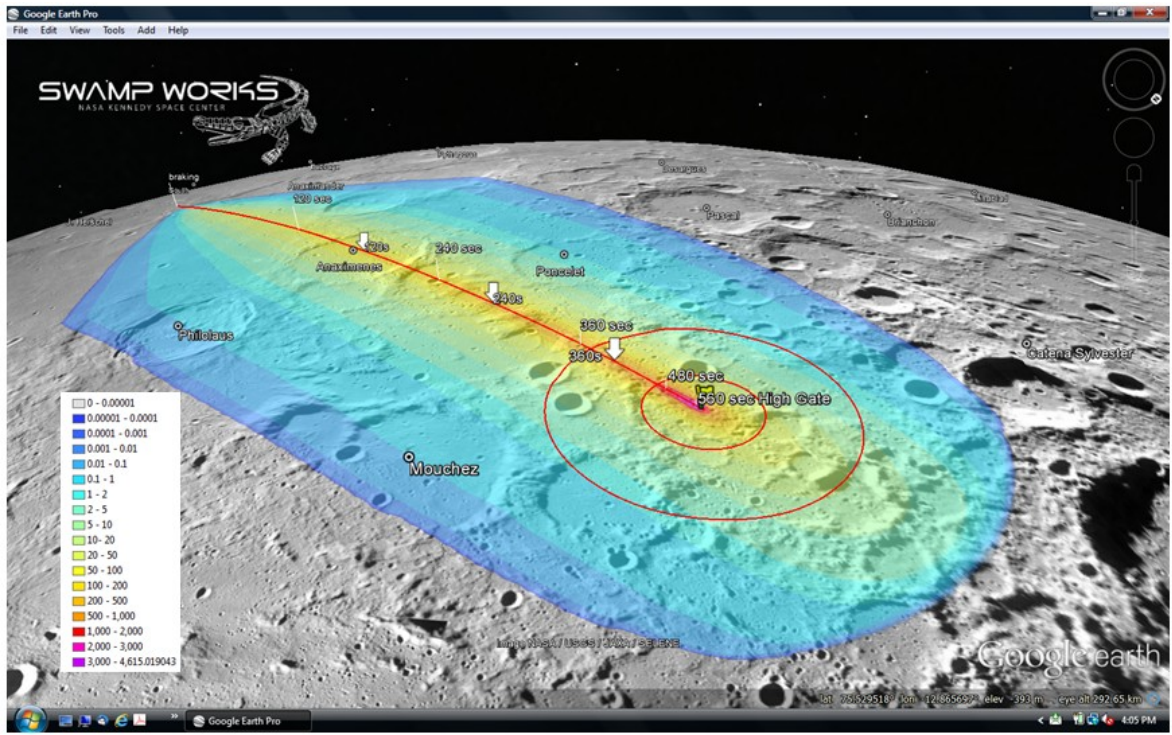

**Figure 18.** Total Lunar deposition from Figure 17 remapped onto Google Moon, without engine exhaust cones.

## SUMMARY AND RECOMMENDATION

The techniques demonstrated in this paper show how deposition maps can be constructed from engine descent plans. These deposition maps make it straightforward to estimate direct exhaust deposition to cold traps. Additional techniques support the calculation of secondary deposition after "hopping", provided that the thermal characteristic temperature at each hop is known. The temperature and number of hops enables an estimate of loss during hopping, which is mostly due to photo-ionization, which has a time constant at 1 AU ~ 1.2 day (Stern, 1999). We are currently researching the characteristic time constant for thermal desorption, but this time constant is likely less than 1 hr which may not be significant to estimates of migration to cold traps.

These results inform potential future missions *where contamination by exhaust gases is not desired*. We recommend that most braking be performed while the braking engine is pointed over the horizon, with exhaust gas velocity >> lunar escape velocity. This will ensure that most exhaust leaves the lunar environment. Once the hypothetical lander has lost sufficient momentum, the braking engine apparatus and excess fuel could be dumped to reduce landing mass. This assumes that the braking engine will not be reused for later ascent. It is also important during any masss dumping operation that any remaining excess fuel does not escape. The lander can then descend with a lower mass (less the descent engine and excess fuel) and minimize exhaust during the final descent to the lunar surface.

**Acknowledgments.** We gratefully acknowledge discussions with Resource Prospector (RP) science team, and Patrick Edward, SWRI.

## REFERENCES

Arnold, J.R. (1979) "Ice in the polar regions," *JGR, 86*, 5659-5668.

Crider, D.H. and R.R. Vondrak (2002) "Hydrogen Migration to the Lunar Poles by Solar Wind Bombardment of the Moon," *Adv Space Res, 30*, No. 8, pp 1869-1874.

Haas, W.R. and S. Price (1984) Atmospheric Dispersion of Hypergolic Liquid Rocket Fuels, *ESL-TR-84-18*.

Metzger, P.T., Smith, J. and Lane, J.E. (2011) "Phenomenology of Soil Erosion by Rocket Exhaust on the Moon and the Mauna Kea Lunar Analog Site," *JGR – Planets, 116:E06005*.

Metzger, P.T., Lane, J.E., Immer, C.D. and Clements, S. (2010) "Cratering and Blowing Soil by Rocket Engines During Lunar Landings," *Lunar Settlements,* ed. by Haym Benaroya (CRC Press), pp. 551-576.

Stern, S.A. (1999) "The lunar atmosphere: History, status, current problems, and context," *Reviews of Geophysics*, *37*(4), 453-491.

Vinti, J.P. (1998) "Orbit Determination from Initial Values," *Orbital and Celestial Mechanics, Volume 177*, Part 1, p 29.